\documentclass{webofc}
\usepackage[varg]{txfonts}   
\usepackage{xcolor}
\usepackage{physics}
\usepackage[capitalise]{cleveref}

\newcommand*{\vv}[1]{\vec{\mkern0mu#1}}

\newcommand{\xT}{\mathbf{x}_T}
\newcommand{\pT}{\mathbf{p}_T}

\newcommand{\qT}{\mathbf{q}_T}

\newcommand{\pTBar}{\bar{\mathbf{p}}_T}
\newcommand{\pVecBar}{\vv{\bar{p}}}
\newcommand{\MBar}{\bar{M}}

\newcommand{\pVec}{\vv{p}}

\newcommand{\wTilde}{\tilde{w}}

\newcommand{\taumin}{{\tau_\text{min}}}
\newcommand{\taumax}{{\tau_\text{max}}}
\newcommand{\dilepton}{{l\bar{l}}}
\newcommand{\entropyconst}{(T\tau^{1/3})_\infty}

\newcommand{\eq}{\text{eq}}

\newcommand{\ideal}{\text{ideal}}

\newcommand{\kompost}{K{\o}MP{\o}ST}

\newcommand{\CIdeal}{\tilde{C}_\gamma^\ideal}

\begin{document}
\title{Pre-equilibrium Photon and Dilepton Production}
%
%

\author{\firstname{Oscar} \lastname{Garcia-Montero}\inst{1}
\and
\firstname{Aleksas} \lastname{Mazeliauskas}\inst{2}
\and
\firstname{Philip} \lastname{Plaschke}\inst{1} \fnsep\thanks{\email{pplaschke@physik.uni-bielefeld.de}}
\and
\firstname{Sören} \lastname{Schlichting}\inst{1}
}

\institute{Fakult\"at f\"ur Physik, Universit\"at Bielefeld\\
D-33615 Bielefeld, Germany
\and
           Institut f\"ur Theoretische Physik, Universit\"at Heidelberg \\
D-69120 Heidelberg, Germany
}

\abstract{%
  We use QCD kinetic theory to compute photon and dilepton production in the chemically equilibrating out-of-equilibrium quark-gluon plasma created in the early stages of high-energy heavy-ion collisions. We derive universal scaling functions for the pre-equilibrium spectra of photons and dileptons. These scaling functions can be used to make realistic predictions for the pre-equilibrium emission and consequently establish the significance of the pre-equilibrium phase  for the production of electromagnetic probes in heavy-ion collisions.
}
\maketitle

\section{Introduction} \label{intro}

Ultra-relativistic heavy-ion collisions produce electromagnetic probes (photons and dileptons) throughout the whole space-time evolution. Both are extraordinary good probes of each stage in such a collision since once they are produced they escape the fireball essentially unmodified and thus provide direct information on their production stage. In this short note we will concentrate on photons and dileptons produced in the quark-gluon plasma (QGP). The QGP, produced shortly after the initial collision, experiences extreme conditions in terms of temperature/energy densities. As such, it is initially far from equilibrium before it will eventually thermalize. With regards to the emission of photons and dileptons, the pre-equilibrium phase of the QGP is often completely neglected or treated in a simplified way. Although there are notable exceptions, a complete leading order (LO) pre-equilibrium evolution together with the production of electromagnetic probes is crucial to understand emission from the QGP. For this, we use QCD kinetic theory~\cite{Arnold:2002zm} in a longitudinally boost invariant and transverse homogeneous system, which provides a smooth transition from early-time evolution to a hydrodynamic description of the plasma~\cite{Kurkela:2018wud}. Within this framework we will compute all LO-processes for quarks and gluons as well as photons and dileptons. Regulations of infrared divergences but also phenomena like the Landau-Pomeranchuk-Migdal effect (LPM-effect) are included into our calculation. In this work we derive universal scaling functions for the pre-equilibrium spectra of photons and dileptons, which only depend on the specific shear viscosity $\eta/s$ and a local energy scale associated with the entropy $\entropyconst^{3} \sim \pqty{\tau s(T)}_\infty \sim \frac{dS}{d^2\xT d\zeta} = \text{\textsl{const}}$, where the entropy density per unit rapidity $\zeta$ becomes constant at times sufficiently late after the collision, s.t. the system is close to local equilibrium, but early enough that transverse expansion can be neglected.

\section{Universal scaling functions for the photon and dilepton spectra} \label{UniversalScalingFunctions}
The production of photons and dileptons in the equilibrating plasma is controlled by two scales: the time scale $\tau_\eq(\xT)$, on which the system approaches a region described by viscous hydrodynamics, as well as the temperature at this time $T(\tau_\eq(\xT)) \equiv T_\eq(\xT)$, which can be further related to the local entropy density of the system $\frac{dS}{d^2\xT d\zeta}$ through
\begin{align} \label{eq:EntropyRel}
    \frac{(2 \pi)^2}{90} \nu_{\rm eff} \tau_{\rm eq} T_{\rm eq}^3 \approx \frac{dS}{d^2\xT d\zeta} \, ,
\end{align}
s.t. the dependence on the local entropy density can be expressed in terms of
\begin{align}
    (T\tau^{{1}/{3}})^{3/2}_{\infty} = \sqrt{\frac{dS}{d^2\xT d\zeta} \bigg/ \left(\frac{(2 \pi)^2}{90} \nu_{\rm eff}\right)}
\end{align}
Here, $\nu_{\rm eff}\approx 32$ is the effective number of bosonic degrees of freedom of the QGP. Earlier studies (\cite{Giacalone:2019ldn} and refs. therein) indicate that the equilibration in pre-equilibrium systems is controlled by a single relaxation rate $\Gamma_\eq = T_\eq/(4\pi\eta/s)$ and hydrodynamization happens on timescales of the order of unity in the dimensionless time variable $\wTilde = \tau_\eq\Gamma_\eq = \tau_\eq T_\eq/(4\pi\eta/s)$, s.t. we find
\begin{align}\label{eq:TauEQTEQ}
    \tau_\eq T_\eq \approx 4 \pi \eta/s \, .
\end{align}
By the use of the entropy relation \cref{eq:EntropyRel} together with \cref{eq:TauEQTEQ}, the equilibrium time $\tau_\eq$ and equilibrium temperature $T_\eq$ are determined by
\begin{subequations}\label{eq:TauEqTeqScaling}
    \begin{align}
        \tau_\eq &\approx \entropyconst^{-3/2} \ (4\pi\eta/s)^{3/2} \, , \label{eq:TauEqScaling} \\
        T_\eq &\approx \entropyconst^{3/2} \ (4\pi\eta/s)^{-1/2} \label{eq:TeqScaling} \, .
    \end{align}
\end{subequations}

\noindent \textit{Scaling function for dileptons.}
In the context of pre-equilibrium dileptons, the invariant mass $\pqty{M=\sqrt{Q^2}}$ spectrum is given by
\begin{align}
    \frac{dN_\dilepton}{d^2\xT MdM dy_q} = \int_{\taumin}^{\taumax} \! \! \! d\tau \tau \int d\zeta \int d^2q_T \frac{dN_\dilepton}{d^4X d^4Q} \, .
\end{align}
By dimensional analysis, the local production rate ${dN_\dilepton}/{d^4X d^4Q}$ will only depend on the ratios $\tau/\tau_{\eq}$ for a (homogeneous and boost invariant system) and $Q^\mu/T_{\eq}$, where the dilepton momentum $Q^\mu$ is a function of invariant mass $M$, transverse momentum $\qT$ and rapidity $y_q-\zeta$ in the local rest-frame of the QGP. By an appropriate scaling in terms of $\tau_\eq$ and $T_\eq$ one can thus determine a dimensionless scaling function $\mathcal{N}_\dilepton$ as
\begin{align}\label{eq:UniversalScalingDileptons}
    \frac{dN_\dilepton}{d^2\xT MdM dy_q} &= \frac{\tau_\eq^2(\xT) T_\eq^2(\xT)}{(4 \pi)^2} \mathcal{N}_\dilepton \pqty{\MBar} = (\eta/s)^2 \mathcal{N}_\dilepton \pqty{\MBar} \, ,
\end{align}
where the rescaled mass is given by $\MBar = {\sqrt{\eta/s}~M}/{(T \tau^{1/3})^{3/2}_{\infty}}$ and the scaling function itself is obtained from
\begin{align}
    \mathcal{N}_\dilepton \pqty{\MBar} = (4 \pi)^2 \int_{\taumin \over \tau_\eq}^{\taumax \over \tau_\eq} \! \! d{\frac{\tau}{\tau_\eq}} ~{\frac{\tau}{\tau_\eq}} \int d\zeta \int d^2\frac{q_T}{T_\eq} ~\frac{dN_\dilepton}{d^4X d^4Q} \pqty{\frac{\tau}{\tau_\eq}, \frac{Q}{T_\eq}} \, .
\end{align}

\vspace{\baselineskip}
\noindent \textit{Scaling function for photons.} In the case of photons, the pre-equilibrium yield can be obtained analogously as
\begin{align}
    \frac{dN_\gamma}{d^2\xT d^2\pT dy_p} = \int_{\taumin}^{\taumax} \! \! \! d\tau \tau \int d\zeta ~E_p ~\frac{dN_\gamma}{d^4X d^3\pVec} \, ,
\end{align}
and in order to derive the associated scaling function, we will follow exactly the same steps as for dileptons. However, the local production rate ${dN_\gamma}/{d^4X d^3\pVec}$ now has dimensions of energy, which will induce another factor of $T_\eq$ to get dimensionless results. One further complication is that the photon rate contains additional factors of the strong coupling, which will be canceled, if we scale the spectrum by a moment of the equilibrium rate defined as
\begin{align}
    \CIdeal = 4 \int_0^\infty d\left(\frac{p}{T}\right) \left(\frac{p}{T}\right)^4 \Bar{C}^\eq_\gamma \left(\frac{p}{T}\right) \, .
\end{align}
By following this logic, we can define the universal scaling function $\mathcal{N_\gamma}$ through
\begin{align}\label{eq:UniversalScalingPhotons}
    \frac{dN_\gamma}{d^2\xT d^2\pT dy_p} &= \frac{\tau_\eq^2(\xT) T_\eq^2(\xT)}{(4\pi)^2} \CIdeal \mathcal{N}_\gamma \pqty{\pTBar} = (\eta/s)^2 ~\CIdeal \mathcal{N}_\gamma \pqty{\pTBar} \, ,
\end{align}
where the rescaled momentum is given by $\pVecBar= {\sqrt{\eta/s}~\pVec}/\entropyconst^{3/2}$ and the explicit form of $\mathcal{N}_\gamma$ is given by
\begin{align}
    \mathcal{N}_\gamma \pqty{\pTBar} = \frac{(4\pi)^2}{\CIdeal} \int_{\taumin \over \tau_\eq}^{\taumax \over \tau_\eq} \! \! d{\frac{\tau}{\tau_\eq}} ~{\frac{\tau}{\tau_\eq}} \int d\zeta ~\frac{E_p}{T_\eq} \frac{1}{T_\eq} \frac{dN_\gamma}{d^4X d^3\pVec} \left(\frac{X}{\tau_\eq}, \frac{\pVec}{T_\eq} \right) \, .
\end{align}

\begin{figure}[t!]
    \centering
    \includegraphics[width=0.48\textwidth]{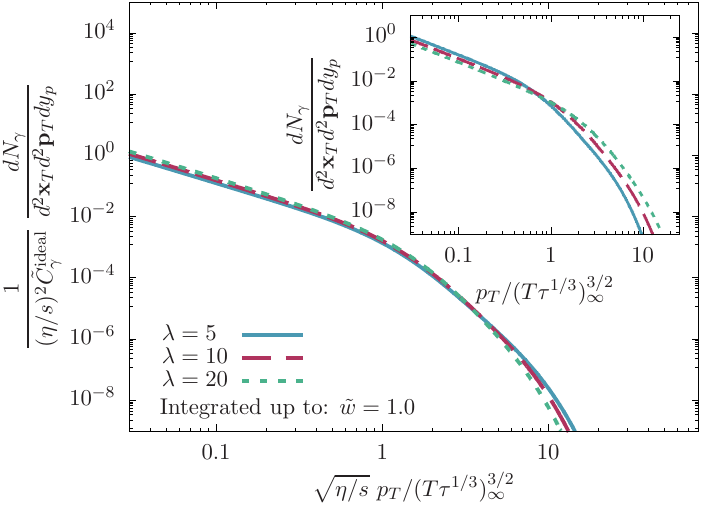}
    \includegraphics[width=0.48\textwidth]{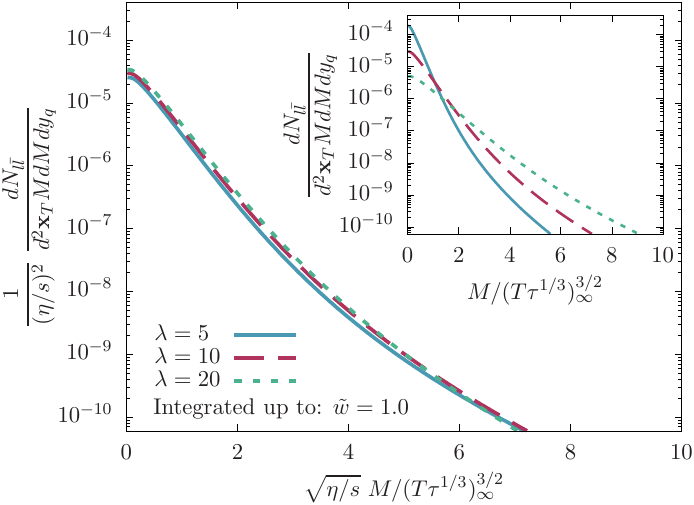}
    \caption{Universal scaling functions for photons (\textit{left}) and dileptons (\textit{right}) as a function of rescaled momentum $\bar{p}_T$ respectively rescaled mass $\bar{M}$. Different colors correspond to different coupling strengths. Each spectrum is integrated up to $\wTilde=1$. The small inset plots show the unscaled spectrum of the respectively particle.}
    \label{fig:ScalingPlots}
\end{figure}


We note that besides the dimensional derivation, it is also possible to derive explicit expressions for $\mathcal{N}_\gamma$ and $\mathcal{N}_\dilepton$, by assuming a universal dependence of the phase-space distribution of quarks and gluons on $T_{\rm eq}$ and $\tau_{\rm eq}$ and we refer to~\cite{Garcia-Montero:2023lrd} for further details for photons. We further note, that in principle it is also possible to derive such scaling functions for other quantities like the transverse mass spectrum for dileptons.

By performing QCD kinetic theory calculations, as described in~\cite{Garcia-Montero:2023lrd}, the proposed scaling is confirmed by our the shown in \cref{fig:ScalingPlots}, which show the time integrated spectra up to $\wTilde=1.0$, where an effective hydrodynamic description of the QGP becomes applicable. 
In~\cref{fig:ScalingPlots} the inset plots shows the unscaled spectra, while the main part shows the corresponding scalings on both axes. Different colors correspond to different couplings $\lambda = 4\pi N_c \alpha_s$. Notably, the spectra for the different couplings get much closer to each other in both cases, where the scaling seems to be slightly better for photons. At larger momentum/large masses, the scaling looses accuracy as these photonts/dileptons are produced at the early instances of time and are hardly modified afterwards. During these times the system is very sensitive to the initial distribution of quarks and gluons. Since the production of quarks from a gluon dominated initial state differs for different couplings, it is expected that the scaling does not hold for very early times ($\wTilde \lesssim 0.15$).

\section{Conclusion and Outlook}
We derived universal scaling functions for pre-equilibrium photon and dilepton production and calculated them in in QCD kinetic theory. Clearly, these scaling functions are powerful tools as they allow  to compute event-by-event pre-equilibrium spectra of photons and dileptons in a very simple way by matching it to realistic values of the specific shear viscosity and of the local energy scale $\entropyconst$ (respectively equilibrium rate $\CIdeal$ for photons). By using these formulae, it is possible to make realistic predictions for the emission from the pre-equilibrium phase of the QGP, such that different sources of photons~\cite{Garcia-Montero:2023lrd} and dileptons can be compared to each other. In order to facilitate the use of our results, we implemented the photon spectrum into the initial state framework \kompost, where the new version (Shiny{\kompost}) and is publicly available under~\cite{KoMPoST}. Under the same source, we will provide Shiny{\kompost} with dileptons in the near future.

Since the $p_T$- and $M$-spectra are rather simple observable for photons and dileptons, a more detailed analysis of differential spectra could help to distinguish between different sources of photon and dilepton production in heavy-ion collisions. Since the universal scaling is obtained on rather general grounds, one can expect such differential observables to follow a similar scaling, and it should in principle also be possible to extend this formalism to other pre-equilibrium processes such as heavy-flavor production, heavy-quark diffusion or jet energy loss, which should be investigated in further studies.

\noindent \textbf{Acknowledgments:} PP and SS acknowledge support by the Deutsche Forschungsgemeinschaft (DFG, German Research Foundation) through the CRC-TR 211 ‘Strong-interaction matter under extreme conditions’-project number 315477589 – TRR 211. 
AM acknowledges support by the DFG through Emmy Noether Programme (project number 496831614)
and CRC 1225 ISOQUANT (project number 27381115).
OGM and SS acknowledge support by the German Bundesministerium für Bildung und Forschung (BMBF) through Grant No. 05P21PBCAA. 

\bibliography{refs}
\clearpage

\end{document}